\documentclass[12pt]{article}
\usepackage{epsfig}

\begin{document}

\title{\bf The width of a chaotic layer}

\author{Ivan~I.~Shevchenko\/\thanks{E-mail:~iis@gao.spb.ru} \\
Pulkovo Observatory of the Russian Academy of Sciences \\
Pulkovskoje ave.~65/1, St.Petersburg 196140, Russia}
\date{}

\maketitle

\begin{center}
Abstract
\end{center}

\noindent A model of nonlinear resonance as a periodically
perturbed pendulum is considered, and a new method of analytical
estimating the width of a chaotic layer near the separatrices of
the resonance is derived for the case of slow perturbation (the
case of adiabatic chaos). The method turns out to be successful
not only in the case of adiabatic chaos, but in the case of
intermediate perturbation frequencies as well.


\section{Introduction}
\label{intro}

The extent of chaotic domains, and, in particular, the width of
chaotic layers, is one of the most important characteristics of
the chaotic motion of Hamiltonian systems. Until now, several
aspects of the problem of analytical estimation of the width of a
chaotic layer were considered in
Refs.~\cite{C78,C79,L91,A96,T98,V04,S04a}. Potentially, the
ability of estimating the extent of chaos in phase space of
Hamiltonian systems has a wide field of applications in physics
and dynamical astronomy. Wisdom et al.~\cite{WPM84} and
Wisdom~\cite{W87} estimated the width of the chaotic layer near
the separatrices of spin-orbit resonances in the rotational
dynamics of planetary satellites and Mercury. Yamagishi~\cite{Y95}
made estimates of the width of the chaotic layer near the magnetic
separatrix in poloidal diverter tokamaks. In these both
applications, Chirikov's approach~\cite{C79} based on the
separatrix map theory was used. Chirikov derived approximate
formulas for the width in the assumption of high-frequency
perturbation of non-linear resonance; however, as follows from
these same formulas, the chaotic layer is exponentially thin with
the ratio of perturbation frequency to the frequency of
small-amplitude phase oscillations on the resonance. This means
that the cases of intermediate and low frequencies of perturbation
are most actual in applications. So, analysis of the problem of
estimation of the width of a chaotic layer in these cases is
definitely necessary.

In this paper, a method of analytical estimation of the width of a
chaotic layer, especially aimed at the case of slow, or adiabatic,
chaos, is proposed. It is based on the theory of separatrix maps.
The nonlinear resonance is modelled by the Hamiltonian of a
perturbed nonlinear pendulum. There are two fundamental
parameters: the ratio of the frequency of perturbation to the
frequency of small-amplitude phase oscillations on the
resonance, and the parameter characterizing strength of the
perturbation.

The applicability of the theory of separatrix maps for description
of the motion near the separatrices of the perturbed nonlinear
resonance in the full range of the relative frequency of
perturbation, including its low values, was discussed and shown to
be legitimate in Ref.~\cite{S00}.

The field of applications of the derived method is rather wide due
to generic character~\cite{C79} of the perturbed pendulum model of
nonlinear resonance. The method can be used in any application
where a separatrix map is derived for description of chaotic
motion. Many of such applications are described, e.g., in
Ref.~\cite{A06}.

Analytical and numerical approaches to measuring the width of a
chaotic layer have different merits and different demerits. The
inherent shortcoming of any analytical approach consists in that
it implies an idealization of the phenomenon, and the estimates
are inherently approximate. The precision of the estimates is hard
to evaluate, due to a number of approximations involved. On the
other hand, the numerical methods are applicable in a rather
narrow range of values of parameters: they cannot be used in the
case of very low relative frequencies of perturbation (due to
limitations on computation time), also in the case of high
relative frequencies of perturbation (because the width of the
chaotic layer is exponentially thin with the perturbation
frequency), and in the case of very small amplitudes of
perturbation, due to limitations on the arithmetic precision.
Therefore only analytical methods can give the global picture.
Their another advantage is that the analytical estimation is easy
and fast, as soon as the theoretical model is shown to be valid.
Finally, the most important advantage, perhaps, is in the physical
insight that the analytical methods provide, making the role of
each parameter clearly visible.

\section{The model of nonlinear resonance and the separatrix map}
\label{mnr}

Under general conditions~\cite{C77,C79,LL92}, a model of nonlinear
resonance is provided by the Hamiltonian of the nonlinear pendulum
with periodic perturbations. A number of problems on non-linear
resonances in mechanics and physics is described by the
Hamiltonian

\vspace{-3mm}

\begin{equation}
H = {{{\cal G} p^2} \over 2} - {\cal F} \cos \varphi +
    a \cos(k \varphi - \tau) + b \cos(k \varphi + \tau)
\label{h}
\end{equation}

\noindent (see, e.g., Ref.~\cite{S00}). The first two terms in
Eq.~(\ref{h}) represent the Hamiltonian $H_0$ of the unperturbed
pendulum; $\varphi$ is the pendulum angle (the resonance phase
angle), $p$ is the momentum. The periodic perturbations are given
by the last two terms; $\tau$ is the phase angle of perturbation:
$\tau = \Omega t + \tau_0$, where $\Omega$ is the perturbation
frequency, and $\tau_0$ is the initial phase of the perturbation.
The quantities ${\cal F}$, ${\cal G}$, $a$, $b$, $k$ are
constants. We assume that ${\cal F} > 0$, ${\cal G} > 0$, $k$ is
integer, and $a = b$. We use the notation $\varepsilon \equiv a /
{\cal F} = b / {\cal F}$ for the relative amplitude of
perturbation.

The so-called separatrix (or ``whisker'') map

\begin{eqnarray}
& & w_{i+1} = w_i - W \sin \tau_i,  \nonumber \\
& & \tau_{i+1} = \tau_i +
                 \lambda \ln {32 \over \vert w_{i+1} \vert}
                 \ \ \ (\mbox{mod } 2 \pi),
\label{sm}
\end{eqnarray}

\noindent written in the present form and explored in
Refs.~\cite{C77,C78,C79} and first introduced in Ref.~\cite{ZF65},
describes the motion in the vicinity of the separatrices of
Hamiltonian~(\ref{h}). The quantity $w$ denotes the relative (with
respect to the unperturbed separatrix value) pendulum energy $w
\equiv {H_0 \over {\cal F}} - 1$, and $\tau$ retains its meaning
of the phase angle of perturbation. The constants $\lambda$ and
$W$ are the two basic parameters, already mentioned in the
Introduction. The parameter $\lambda$ is the ratio of $\Omega$,
the perturbation frequency, to $\omega_0 = ({\cal F} {\cal G})^{1/2}$,
the frequency of the small-amplitude pendulum oscillations. The
parameter $W$ in the case of $k=1$ and $a = b$ has the
form~\cite{S98}:

\vspace{-3mm}

\begin{equation}
W = \varepsilon \lambda \left( A_2(\lambda) + A_2(-\lambda) \right) =
{4 \pi \varepsilon \lambda^2 \over \sinh {\pi \lambda \over 2}}.
\label{W1}
\end{equation}

\noindent Here $A_2(\lambda) = 4 \pi \lambda {\exp({{\pi \lambda}
/ 2}) \over \sinh (\pi \lambda)}$ is the value of the
Melnikov--Arnold integral as defined in Ref.~\cite{C79}.
Formula~(\ref{W1}) differs from that given in
Refs.~\cite{C79,LL92} by the term $A_2(-\lambda)$, which is small
for $\lambda \gg 1$. However, its contribution is significant for
$\lambda$ small~\cite{S98}, i.e., in the case of adiabatic chaos.
Expression~(\ref{W1}) for the parameter $W$ needs to be modified
at very high relative frequencies of perturbation (see
Refs.~\cite{G97,T98}). Analytical expressions for $W$ at different
values of $k$ are given in Refs.~\cite{C79,S00} and at arbitrary
$a$, $b$ in Ref.~\cite{S00}.

The accuracy of separatrix map~(\ref{sm}) in describing the
behaviour of original system~(\ref{h}) can be estimated by the
order of magnitude as $\sim \varepsilon$ (see
Refs.~\cite{T98,A06}). Measurement of the chaotic layer width
allows one to estimate the accuracy directly, as demonstrated
below in Section~\ref{aene}.

Note that the expression for the increment of the phase $\tau$ in
map~(\ref{sm}) is a rough approximation. It is valid for a low
strength of perturbation, i.e., at $w \ll 1$. According to
Refs.~\cite{S98,S99}, one can improve the accuracy of the map by
means of replacing the logarithmic approximation of the phase
increment by the original expression through the elliptic
integrals. For the sake of brevity we do not explore the
advantages of this improvement further in estimating the width.
This can be straightforwardly accomplished if one needs to improve
precision of estimating the width at increasing the magnitude of
perturbation.

One iteration of map~(\ref{sm}) corresponds to one period of the
pendulum rotation or a half-period of its libration. The motion of
system~(\ref{h}) is mapped by Eqs.~(\ref{sm})
asynchronously~\cite{S98}: the relative energy variable $w$ is
taken at $\varphi = \pm \pi$, while the perturbation phase $\tau$
is taken at $\varphi = 0$. The desynchronization can be removed by
a special procedure~\cite{S98,S00}. The synchronized separatrix
map gives correct representation of the sections of the phase
space of the near-separatrix motion both at high and low
perturbation frequencies; this was found in Ref.~\cite{S00} by
direct comparison of phase portraits of the separatrix map to the
corresponding sections obtained by numerical integration of the
original systems. This testifies good performance of both the
separatrix map theory and the Melnikov theory (that describes the
splitting of the separatrices).

The asymptotic expression for $W$ that ensues from Eq.~(\ref{W1})
at $\lambda \sim 0$ is $W \approx 8 \varepsilon \lambda$. A good
correspondence of this expression to the actual amplitude of the
separatrix map derived numerically by integration of the original
system was found in Ref.~\cite{V04}.

An equivalent form of Eqs.~(\ref{sm}), used, e.g., in
Refs.~\cite{CS84,S98}, is

\vspace{-3mm}

\begin{eqnarray}
     y_{i+1} &=& y_i + \sin x_i, \nonumber \\
     x_{i+1} &=& x_i - \lambda \ln \vert y_{i+1} \vert + c
                   \ \ \ (\mbox{mod } 2 \pi),
\label{sm1}
\end{eqnarray}

\noindent where $y = w / W$, $x = \tau + \pi$; and

\vspace{-3mm}

\begin{equation}
c = \lambda \ln {32 \over \vert W \vert}.
\label{c}
\end{equation}

\noindent Note that while the second line of Eqs.~(\ref{sm1}) is
taken modulo $2 \pi$, the quantity $c$ given by Eq.~(\ref{c}) is
taken modulo $2 \pi$ in the following analytical treatment only
when explicitly stated.

\section{The case of the least perturbed border}
\label{clpb}

We obtain the dependence of the half-width $y_b$ of the main
chaotic layer of the separatrix map in a numerical experiment with
Eqs.~(\ref{sm1}). The border value $y_b$, corresponding to the
maximum energy deviation (from the unperturbed separatrix) of a
chaotic trajectory inside the layer, is determined as the maximum
of $|y_i|$ obtained during computation of a single chaotic
trajectory.
At each step (equal to $0.05$) in the segment $\lambda \in [0,
10]$, the values of $y_b$ are computed for 100 values of $c$
equally spaced in the interval $[0, 2 \pi)$. The number of
iterations for each trajectory is $n_{it} = 10^8$. This has been
checked to be sufficient to saturate the computed values of $y_b$.
Following the approach of Ref.~\cite{S04a}, at each step in
$\lambda$ we find the value of $c$ corresponding to the minimum
$y_b$ (the case of the least perturbed border), and plot the value
of $y_b$ corresponding to this case; such $y_b$ value is denoted
in what follows as $y_{lb}$.

The initial fragment of thus obtained ``$\lambda$--$y_{lb}$''
relation is plotted in Fig.~1. The observed dependence apparently
follows the piecewise linear law

\vspace{-3mm}

\begin{equation}
y_{lb} \approx \cases{
1, & if $0 \le \lambda \le 1-a$, \cr
\lambda + a, & if $\lambda > 1-a$,
}
\label{ylbmax}
\end{equation}

\noindent where $a \approx 1/2$ apparently.

The linear fit $y_{lb}(\lambda) = {a + b \lambda}$ of the observed
dependence at $\lambda \in [0.5, 10]$ gives the following values
for the coefficients and their standard errors: $a = 0.5351 \pm
0.0041$, $b = 1.0059 \pm 0.0007$. The $y_{lb}(\lambda)$ dependence
represents the sum of two addends. The first addend, $a$, is the
half-amplitude of the last rotational invariant curve of the
standard map at the critical value of the stochasticity parameter,
because the separatrix map at high values of $\lambda$ is locally
(in $y$) approximated by the standard map. The theoretical value
of this addend, which can be found by direct computation with the
standard map, is $\approx 0.508$. The second addend, $b \lambda$,
is the border value of $y$ averaged over $x \in [0, 2 \pi]$. The
theoretical value of this addend approximately equals $\lambda$,
see Ref.~\cite{C79}. So, the deviations of the observed values of
$a$ and $b$ from their theoretical values are rather small,
$0.027$ and $0.006$ respectively.

One can eliminate the constant component, arising due to the fact
that the border curve is not a horizontal straight line, by
subtracting the half-amplitude of the border curve. The resulting
value coincides with the value of $y$ time-averaged on the
trajectory following the boundary curve. We designate this value
by $\bar y_{lb}$.  At $\lambda \ge {1 \over 2}$, one has $\bar
y_{lb} \approx \lambda$ apparently. At $\lambda \le {1 \over 2}$,
$\bar y_{lb}$ equals one half of the maximum $y$ value; this
follows from the form of the border curve derived in the next
Section. So, one obtains for the time-averaged half-width

\vspace{-3mm}

\begin{equation}
\bar y_{lb} \approx
\cases{
{1 \over 2}, & if $0 \le \lambda \le {1 \over 2}$, \cr
\lambda, & if $\lambda > {1 \over 2}$.
}
\label{ylb}
\end{equation}

\noindent The clear break at the point $\lambda \approx 1/2$
manifests a physical distinction between two types of dynamics,
namely, {\it slow} and {\it fast} chaos. While in the case of slow
chaos the chaotization of motion can be physically explained by
the sporadic encounters with the singular line $y=0$ (such
mechanism was originally evoked in an analysis of the so-called
``relativistic'' map in Ref.~\cite{ZSUC91}), in the case of fast
chaos the natural traditional explanation consists in the
phenomenon of resonance overlapping~\cite{C79}.

Approximating the separatrix map by the standard one locally in
$y$, Chirikov~\cite{C79} derived the linear $\lambda$ dependence
for the width of the chaotic layer at $\lambda \gg 1$. By means of
a rigourous mathematical argument, Ahn et al.~\cite{A96} were able
to set a lower bound on the width of the chaotic layer. In our
notation, this lower bound is given by the formula $y_b >
\frac{3}{4} \lambda$ (see Eq.~(5.8) in Ref.~\cite{A96}).
Eq.~(\ref{ylb}) is in accord with these theoretical findings.

The formula for the time-averaged half-width in the original
energy variable $w$ is

\vspace{-3mm}

\begin{equation}
\overline w_{lb} = \vert W \vert \bar y_{lb} = {4 \pi \vert
\varepsilon \vert \lambda^2 \bar y_{lb} \over \sinh {\pi \lambda
\over 2}}, \label{wlb}
\end{equation}

\noindent where $\bar y_{lb}$ is given by Eq.~(\ref{ylb}).
Formula~(\ref{wlb}) is valid for any frequency of perturbation in
system~(\ref{h}) with $k=1$, $a = b$, provided that the separatrix
map correctly describes the behaviour of the original system, and
the analytical representation of $W$ is correct. At very high
frequencies of perturbation, expression~(\ref{W1}) for the
parameter $W$ needs correction, as noted above. At low frequencies
of perturbation, formula~(\ref{wlb}) demonstrates that the chaotic
layer width, expressed in $w$, decreases linearly with $\lambda$.
This is due to the fact that we consider the case of the least
perturbed border, and therefore $\varepsilon$ is not fixed.

If one fixes $\varepsilon$, the low frequency limit of the width
is a non-zero constant. This is a trivial consequence of the usual
``slowly pulsating separatrix'' approach considered e.g.\ in
Refs.~\cite{N86,BC89}. Indeed, Hamiltonian~(\ref{h}) with $k=1$,
$a=b$ is naturally set in the form of that of a pendulum with
modulated frequency of small-amplitude oscillations:

\vspace{-3mm}

\begin{equation}
H = {{{\cal G} p^2} \over 2} - ({\cal F} - 2 a \cos \tau ) \cos
\varphi
\label{hmp}
\end{equation}

\noindent (see, e.g., Ref.~\cite{C79}). Considering the relative
full energy $w_H \equiv {H \over {\cal F}} - 1$ instead of $w =
{H_0 \over {\cal F}} - 1$, one can get a simple heuristic estimate
for the width of the chaotic layer in the limit $\lambda \to 0$.
Indeed, as follows from representation~(\ref{hmp}), the energy
$w_H$ on the slowly pulsating separatrix varies from $-2
\varepsilon$ to $2 \varepsilon$; so, the half-width of the layer
is equal to $2 \varepsilon$. This quantity, however, includes the
amplitude of the chaotic layer bending (the effect, described
in~\cite{S00}). Let us note again that this ``zero $\lambda$''
half-width estimate is made in the units of the relative full
energy $w_H$. Below we shall see that, in the $w$ units, the
``zero $\lambda$'' half-width incorporating bending is equal to $4
\varepsilon$, and the ``zero $\lambda$'' sheer half-width is equal
to $\varepsilon$ (for the same case of $k=1$ and $a=b$, of course;
in other cases the widths are generally different).

\section{The general formulas at slow perturbation}
\label{gfsp}

Let us assume that $\lambda \ll 1$. In this case the diffusion
across the layer is slow, and on a short time interval the phase
point follows close to some current curve. We call this curve
guiding. Let us derive an analytical expression for the guiding
curve with an irrational winding number far enough from the main
rationals. We approximate the winding number $Q$ by the
rationals $m/n$. Thus $c \approx 2 \pi m/n$. Noticing that at an
iteration $n$ of the map the phase point hits in a small
neighbourhood of the starting point, one obtains for the
derivative

\vspace{-3mm}

\begin{eqnarray}
\frac{dy}{dx} &=& \frac{1}{nc - 2\pi m} \sum_{k=0}^{n-1} \sin(x+kc) = \nonumber \\
&=& \frac{1}{nc - 2\pi m} \sin {n c \over 2} \mbox{cosec } {c
\over 2} \sin \left(x + {n-1 \over 2} c \right)
\label{dydx}
\end{eqnarray}

\noindent (the second equality follows from formula (1.341.1) in
Ref.~\cite{GR62}). Integrating and passing to the limit $n \to
\infty$, one obtains finally

\vspace{-3mm}

\begin{equation}
y = -\frac{1}{2} \mbox{cosec } \frac{c}{2} \cos \left(x -
\frac{c}{2} \right) + {\cal C}, \label{yC}
\end{equation}

\noindent where ${\cal C}$ is an arbitrary constant of
integration. Description~(\ref{yC}) inside the chaotic layer is
approximately valid on short intervals of time, for which the slow
diffusion across the layer can be neglected.

The motion is chaotic only when curve~(\ref{yC}) crosses the line
$y=0$ which is the inverse image of the singular curve  $y= -
\sin x$. Hence the half-width of the chaotic layer is

\vspace{-3mm}

\begin{equation}
y_b = \left| \mbox{cosec} \frac{c}{2} \right|, \label{yb}
\end{equation}

\noindent or, equivalently, the half-width in the original energy
variable $w$ is

\vspace{-3mm}

\begin{equation}
w_b = \left| W \mbox{cosec} \left( \frac{\lambda}{2} \ln {32 \over
\vert W \vert} \right) \right|. \label{wb}
\end{equation}

\noindent The quantities $y_b$ and $w_b$ are the {\it maximum}
relative energy deviations inside the layer; the {\it
time-averaged} half-widths $\bar y_b$ and $\overline w_b$, as
follows from the geometry of the boundary curve, Eq.~(\ref{yC}),
are two times less.

A different approach for passing from map~(\ref{sm}) to a
differential equation in the case of slow chaos was used in
Ref.~\cite{V04}. That approach assumes that the increments in the
map~(\ref{sm}) variables are small at {\it each} iteration of the
map. From our representation~(\ref{sm1}) of the separatrix map it
is clear that such condition can be satisfied only when $y \gg 1$,
and therefore $y_b \gg 1$; from Eq.~(\ref{yb}) it follows then
that the value of $c$ should be close to the ``main resonance''
value $\approx 0 \mbox{ mod } 2 \pi$. On this condition the
increment in $x$ at each iteration of Eqs.~(\ref{sm1}) is small as
well.

What are the restrictions for the validity of our
approximation~(\ref{yC})? Since, in deriving the increment in $x$,
we have neglected the term $\lambda \ln \vert y_{i+1} \vert$ in
the second equation of Eqs.~(\ref{sm1}), the inequality
$c \mbox{ mod } 2 \pi \gg \lambda \ln \vert y_b \vert$ should hold, i.e.,

\vspace{-3mm}

\begin{equation}
c \mbox{ mod } 2 \pi \gg \lambda \ln \left| \mbox{cosec}
\frac{c}{2} \right|. \label{ctr}
\end{equation}

\noindent So, since $\lambda$ is small, the value of $c$ should be
far enough from the main resonance $c \approx 0 \mbox{ mod } 2
\pi$. Besides, the value of $c$ should not correspond to other
resonances, the role of which, however, is much less.

Expression~(\ref{yb}) is compared to the numerical experimental
data for $\lambda = 0.1$ in Fig.~2. The maximum relative energy
deviations have been found for the chaotic trajectories of
Eqs.~(\ref{sm1}) in the interval $c \in [0, 2 \pi)$ with the step
in $c$ equal to $0.001$; each trajectory has been computed on the
time interval $n_{it} = 10^8$. This has been checked to be
sufficient to saturate the computed values of $y_b$. It is evident
that the theoretical curve follows closely the numerical data at
all values of $c$ except the resonant ones, where discontinuities
are observed. The latter ones are conditioned by appearance of
regular islands inside the chaotic layer at the resonant values of
$c$. These perturbations are similar to those in the behaviour of
the standard map, for which the small local peaks in the
dependence of the maximum Lyapunov exponent on the stochasticity
parameter are conditioned by the local depressions in the measure
of the chaotic component of phase space, due to appearance of
regular islands~\cite{S04c}.

The deviations of the theory from numerics are most visible near
the main resonance, i.e., at $c$ near $0 \mbox{ mod } 2 \pi$. If
$\lambda$ and $c$ are close to zero, the relative increments of
$w$ and $\tau$ in Eqs.~(\ref{sm}) are small. Then, as mentioned
above, approach~\cite{V04} for passing from map~(\ref{sm}) to a
differential equation is applicable, and Eqs.~(\ref{sm}) can be
approximated by the differential equation

\vspace{-3mm}

\begin{equation}
\frac{dw}{d\tau} = - \frac{W \sin \tau}{\lambda \ln {32 \over
\vert w \vert}}
\label{dwdtau}
\end{equation}

\noindent analogous to that derived in Ref.~\cite{V04}, except
that we use homogeneous variables here and take into account the
condition on $c$. Similarly to Eq.~(\ref{dydx}),
Eq.~(\ref{dwdtau}) describes a guiding curve with an arbitrary
constant of integration ${\cal C}$:

\vspace{-1mm}

\begin{equation}
w \ln {32 e \over \vert w \vert} = \frac{W}{\lambda} (\cos \tau +
{\cal C}). \label{wtau}
\end{equation}

\noindent As in derivation of Eq.~(\ref{yb}), we note that the
motion is chaotic only when curve~(\ref{wtau}) crosses the line
$w=0$. Then the constant of integration for the boundary curves of
the chaotic layer is ${\cal C} = \pm 1$, and the equation for the
half-width $w_b$ of the chaotic layer is

\vspace{-1mm}

\begin{equation}
w_b \ln {32 e \over w_b} = \frac{2 | W |}{\lambda}.
\label{wmr}
\end{equation}

\noindent The approximation of $W$, which should be substituted
here in the case of $k = 1$, follows from Eq.~(\ref{W1}): if
$\lambda \ll 1$, one has $W \approx 8 \varepsilon \lambda$, then

\vspace{-1mm}

\begin{equation}
w_b \ln {32 e \over w_b} = 16 | \varepsilon |. \label{wmr1}
\end{equation}

\noindent So, at small values of $\lambda$ the half-width $w_b$
depends solely on $\varepsilon$; the $\lambda$ dependence expires.
For a different value of $k$, the formula for $W$ and its
approximation are different (see Refs.~\cite{S00,V04}), but the
$\lambda$ dependence in Eq.~(\ref{wmr}) expires all the same.
Eq.~(\ref{wmr1}) (or (\ref{wmr})) is easily solved numerically by
iterations.

In summary, formulas~(\ref{yb}) and (\ref{wb}) for the half-width
of the chaotic layer are applicable in the case of generic values
of the $c$ parameter (excluding the main resonance case), and
Eqs.~(\ref{wmr}) and (\ref{wmr1}) are applicable in the main
resonance case $c \approx 0 \mbox{ mod } 2 \pi$. In the first case
the chaotic layer width depends on both $\lambda$ and
$\varepsilon$, while in the second case the dependence on
$\lambda$ expires and the width depends solely on $\varepsilon$.

\section{Analytical estimates versus numerical experiment}
\label{aene}

To check the theory, the width of the chaotic layer near the
separatrices of Hamiltonian~(\ref{h}) has been directly computed.
The integration of the equations of motion has been performed by
the integrator by Hairer et al.~\cite{HNV87}. It is an explicit
8th order Runge--Kutta method due to Dormand and Prince, with the
step size control.

At each value of $\lambda$ the half-width has been measured by two
methods. The first one was proposed in Refs.~\cite{C78,C79} and
developed and extensively used in Ref.~\cite{V04}. It is based on
calculation of the minimum period $T_{min}$ of the motion in the
chaotic layer. The half-width is determined by the
formula~\cite{C78,C79,V04}:

\vspace{-3mm}

\begin{equation}
w_b =  32 \exp(- \omega_0 T_{min}).
\label{wb1}
\end{equation}

\noindent The minimum period corresponds to the maximum energy
deviation from the unperturbed separatrix value. This formula
directly follows from the second line of Eqs.~(\ref{sm}).

The second method consists in the direct continuous
measuring of the relative energy deviation from the unperturbed
separatrix $w = {H_0 \over {\cal F}} - 1$ in the course of
integration, and fixing the extremum one.

The integration time interval has been chosen to be $10^3$, in the
units of periods of perturbation. Each unit is divided in $10^5$
equal segments; the trajectories have been output at the end of
each segment, to provide, with such time resolution, the
calculation of the time period and the relative energy deviation
of the motion. Further increasing the integration time interval or
decreasing the length of the segments have been checked to leave
the estimates of the width unchanged within 3--4 significant
digits; i.e., the estimates are saturated enough.

We set $k=1$, $a=b$, $\varepsilon = 10^{-5}$. The results of
calculation of $w_b$ by the first method are shown in Fig.~3, by
the second method in Fig.~4. Logarithmic horizontal scale is used;
decimal logarithms are implied by ``$\log$'' throughout this
paper. The experimental values are shown by dots. The theoretical
dependence, given by Eq.~(\ref{wb}), is shown in both figures by
solid curve. General qualitative agreement is observed between the
theory and the experimental data up to $\lambda \approx 1$, i.e.,
up to the intermediate values of the frequency of perturbation;
even the sharp variations are in qualitative accord. Theoretical
dependence~(\ref{wb}) visually follows the experimental one even
at high values of $\lambda$. This is solely a visual effect:
deviations in locations of resonance peaks increase with
increasing $\lambda$. Eq.~(\ref{wb}) is applicable at low and
intermediate relative frequencies of perturbation; at high
frequencies a different formula, namely Eq.~(\ref{wlb}), should be
used. This latter formula describes the behaviour averaged over
$c$.

Before analyzing the quantitative agreement of the theory and the
experimental data, let us consider a notable feature of the
observed dependence: the sharp peaks. They are conditioned by the
process of encountering the main resonances $c \approx 2 \pi m$,
$m = 1, 2, \dots$, as $\lambda$ increases. According to
Eqs.~(\ref{yb},~\ref{wb}), at such values of $c$ the width goes to
infinity; the real width is finite, of course. The approximate
location of the peaks in $\lambda$ is determined by the equation

\vspace{-3mm}

\begin{equation}
c = \lambda \ln {4 \over \vert \varepsilon \vert \lambda} = 2 \pi
m. \label{peaks}
\end{equation}

\noindent The location is practically insensitive to the value of
$k$. If $\vert \varepsilon \vert \ll \lambda$, the location can be
estimated by the formula $\lambda_m \approx - 2 \pi m / \ln
\varepsilon$ very approximately; i.e., on decreasing $\varepsilon$
the peaks move slowly to the left.

The abscissas $\lambda_m$ of the peaks in Figs.~3 and 4 are all
greater than $0.5$, i.e., the peaks are situated in the domain of
fast chaos. In such circumstances, the ``tangency
condition''~\cite{S98} for the marginal integer resonances should
provide better precision for estimating the location of the peaks.
An opportunity of sporadic strong variations of the relative
energy $w$ in the motion in the chaotic layer at $\lambda
> 1$ depends on the structure of the border of the chaotic
layer~\cite{C79,S98}. Excursions to high values of the relative
energy $w$ become possible, when, with variation of a parameter,
the border of the chaotic layer starts to overlap with the narrow
chaotic layer near the separatrices of an integer resonance, i.e.,
a heteroclinic connection emerges between them. Since the first
layer is very narrow, this phenomenon can be approximately
described as ``tangency'' of the unperturbed separatrix of the
marginal resonance and the border of the main chaotic layer. The
equation for the tangency condition~\cite{S98} is

\vspace{-1mm}

\begin{equation}
W = W_t^{(m)}(\lambda), \label{tc}
\end{equation}

\noindent where $W$ at $k=1$, $a=b$, is given by Eq.~(\ref{W1})
and

\vspace{-3mm}

\begin{equation}
W_t^{(m)}(\lambda) = {32 \over \lambda^3}
    \left( \left( 1 + \lambda^2 \right)^{1/2} - 1 \right)^2
    \exp \left( - {2 \pi m \over \lambda} \right).
\label{Wt}
\end{equation}

\noindent Eq.~(\ref{tc}), solved numerically, does provide good
precision: the calculated abscissas of the first five peaks $\log
\lambda_m$ ($m=1, 2, 3, 4, 5$) deviate from the observed in the
numerical experiment not greater than by $0.02$. The deviations
are all positive. The accuracy of the observed values is
determined by the horizontal resolution of the experimental plots,
which is $\log 1.01 \approx 0.004$.

As expected, Eq.~(\ref{peaks}) is less successful: it predicts
$\log \lambda_m = -0.34$, $-0.01$, $0.18$, $0.32$, $0.42$ ($m=1,
2, \dots, 5$) versus the observed values $-0.28$, $0.02$, $0.19$,
$0.30$, $0.39$, i.e., the deviation in absolute value is not
greater than $0.06$.

The tangency condition can be employed for analytical estimating
the height of the peaks; see Eq.~(10) in Ref.~\cite{S98}.

While the main resonances with $m = 1, 2, \dots$ manifest
themselves in the peaks, that with $m=0$ results in the asymptotic
horizontal plateau at $\lambda \to 0$. This plateau was found and
discussed in Ref.~\cite{V04}, and an approximate heuristic formula
was proposed for its asymptotic height: $w_b/\varepsilon \approx
0.22 \cdot 8 = 1.76$ for $k=1$ (see Eq.~(14) of that paper; a
misprint (missing $\varepsilon$) is corrected here). This estimate
differs from our data less than by a factor of 2.

Our experimental $w_b/\varepsilon$ value at $\lambda = 0.01$ is
equal to $1.00585$. Numerical solving Eq.~(\ref{wmr1}) gives
$w_b/\varepsilon = 1.00142$, i.e., a value perfectly close to the
experimental one; the difference is only $0.4$\%.
Formula~(\ref{wb}), giving $w_b/\varepsilon = 0.915$, is also
quite successful, notwithstanding the fact that this part of
dependence is beyond the scope of its applicability ($c$ is close
to zero).

The reason for the plateau emerging at small values of $\lambda$
is clear: if one fixes $\varepsilon$, no matter how small this
fixed value is, on decreasing $\lambda$ the value of $c = \lambda
\ln {32 \over \vert W \vert}$ also decreases and inevitably finds
itself near the main resonance $c \approx 0$. The transition point
to the main resonance domain is determined by the abscissa of the
point of intersection between the curve given by
formula~(\ref{wb}) (this curve goes asymptotically to zero, though
slowly) and the horizontal line defined by Eq.~(\ref{wmr}).

In Fig.~3, this transition point is situated at $\log \lambda
\approx -1.44$. To the left of this point, the deviations of the
observed $w_b$ data from the theoretical plateau level $1.00142$
do not exceed 2\%. To the right of the point, the deviations of
the solid curve (given by formula~(\ref{wb})) from the observed
data is less than 10\% in absolute value typically, except that at
narrow resonant intervals in $\lambda$ (fractional resonances
manifest themselves as small peaks) the deviations rise up to
about 30\%. This agreement continues up to $\log \lambda \approx
-0.4$. Then the first integer resonance comes into play, and, due
to incomplete correspondence in real and theoretical locations of
the peak, the deviation rises sharply. At greater values of
$\lambda$ the agreement is solely qualitative.

The ``$\lambda$--$w_b$'' dependence constructed by the second
method (in Fig.~4) at small values of $\lambda$ lies notably
higher than the experimental data. This reveals an interesting
phenomenon of the chaotic layer bending, described in
Ref.~\cite{S00}. This strictly geometrical phenomenon is absent in
the previous graph, because it averages out in that case.
According to Ref.~\cite{S00}, the relative energetic amplitude of
bending at $k=1$ in the limit $\lambda \to 0$ at the section of
phase space $\varphi = 0 \mbox{ mod } 2 \pi$ is equal to $4
\varepsilon$. The experimentally observed value of
$w_b/\varepsilon$ at $\lambda = 0.01$ is equal to $4.00158$, in
perfect agreement with this theoretical prediction: the deviation
is only $0.04$\%.

One should emphasize that the theoretical value of bending refers
to the section $\varphi = 0 \mbox{ mod } 2 \pi$; at any other
value of $\varphi \mbox{ mod } 2 \pi$ the bending is, generally
speaking, different. Our experimental procedure gives the value of
the maximum bending. In the considered case ($k=1$) both values
coincide, i.e., the maximum bending is achieved at $\varphi = 0
\mbox{ mod } 2 \pi$.

The numerical data considered above refer all to the case $k=1$.
The cases of different values of $k$ can be considered in a
similar way. To obtain the theoretical dependence
``$\lambda$--$w_b$'', one should simply substitute the relevant
expression for $W$ in Eq.~(\ref{wb}). E.g., in the case $k=2$, $a
= b$, the expression is

\vspace{-1mm}

\begin{equation}
W = {8 \pi \varepsilon \lambda^2 (\lambda^2 - 2) \over 3 \sinh
{\pi \lambda \over 2}}
\label{W2}
\end{equation}

\noindent \cite{S00,V04}. In Fig.~5, the dependences
``$\lambda$--$w_b$'' for this case are shown: the theoretical one
by solid curve, while those obtained by direct integration of
system~(\ref{h}) are represented by dots. The visual agreement
between the theory and numerical experiment at low and
intermediate frequencies of perturbation is as good as in the case
$k=1$. The theoretical plateau level of $w_b/\varepsilon$ in the
case of $k=2$, given by Eq.~(\ref{wmr}), is $1.36140$. The
characteristics of quantitative agreement in the full range of
$\lambda$ are the same as in the case of $k=1$ (see above), only
the location of the transition point, $\log \lambda \approx
-1.42$, is slightly different.

Let us clarify what is the role of resonances by means of
constructing phase portraits of the motion. The main (integer)
resonances $m = 0, 1, 2, \dots$ result in stretching the layer in
the $y$ direction; the motion in these resonances is quite simple:
it follows the guiding curves~(\ref{wtau}). The role of fractional
resonances is more intricate. The phase portrait of separatrix
map~(\ref{sm1}) for the case of the fractional resonance with
winding number $Q = 4/5$ is shown in Fig.~6a; $\lambda = 0.01$ and
$c = 5.0189 \approx 2 \pi Q$. Only the chaotic component is shown.
The choice of the value of $c$ corresponds to the minimum measure
of the chaotic component inside the borders of the layer. The
phase portrait has been obtained by iterating Eqs.~(\ref{sm1})
$10^7$ times. Further increasing $n_{it}$ does not give any
detectable increase of chaotic component. Instead of visualizing
each iteration (this would produce a file of incredible volume),
Fig.~6a represents a ``rasterized'' phase portrait: it is
comprised by the set of pixels explored by a chaotic trajectory on
the grid of $400 \times 400$ pixels.

The ``porous'' structure of the layer at resonance is clearly seen
in the Figure. The main pattern is formed by 5 curves of
sinusoidal form, embedded in broad strips of generic chaos. These
curves are nothing but the singular curve $y = - \sin x$ and its
four consecutive images. The figure demonstrates how the resonant
structure with large amount of inner regular component is formed:
in the case of resonance with winding number equal to the rational
$p/q$, the $q+1$th image of the singular curve $y = - \sin x$
coincides with the initial singular curve exactly; so, the strips
of generic chaos in the neighbourhood of the singular curves are
not blurred.

A small positive or negative shift (by $\approx 0.002$) in $c$
establishes complete visual ergodicity of the motion inside the
layer, i.e., upon the shift, the layer in Fig.~6a would be just a
black band. However, the proximity of resonance influences the
form of the borders of the layer. On further shifting $c$ away
from the resonance, this influence diminishes, and the borders
become closer and closer in their form to the guiding
curve~(\ref{yC}). In Fig.~6b, the value of the $c$ parameter is
shifted from the resonant case by $0.01$ (i.e., $c = 5.0289$). No
regular islands are seen inside the layer. This visual impression
is deceptive: as follows from our numerical experiments, very
small islands can always be found by implementing special
numerical techniques, such as computation of Lyapunov exponents
for a set of trajectories on a fine grid of initial data. So, no
complete ergodicity of motion inside the layer is achieved in
reality. This interesting phenomenon will be discussed elsewhere.

The fractional resonances are barely seen (as small peaks) in
Figs.~3, 4 and 5. On decreasing $\lambda$, no matter what the
value of $\varepsilon$ is, one reaches the plateau corresponding
to the main resonance $m = 0$. Instead, if $\lambda$ is fixed and
$\varepsilon$ is decreased, one finds much more intricate
behaviour. According to formula~(\ref{c}), no matter how small the
value of $\lambda$ is, one can achieve any value of $c$ by
diminishing $\varepsilon$. In other words, all the set of
resonances is traversed once and once again, if one decreases
$\varepsilon$ steadily. Due to logarithmic dependence on
$\varepsilon$, at small values of $\lambda$ (already at $\lambda =
0.01$) the encounters with prominent resonances take place at
microscopic values of $\varepsilon$. Setting $c = 2 \pi (Q + m)$
(where $Q$ is taken modulo 1, and $m = 0, 1, 2, \dots$) and
rearranging Eq.~(\ref{c}), one has for the resonant value of
$\varepsilon$:

\vspace{-1mm}

\begin{equation}
\varepsilon_{res} =\frac{4}{\lambda} \exp \left( - \frac{2 \pi (Q
+ m)}{\lambda} \right).
\label{epsres}
\end{equation}

\noindent If $\lambda = 0.01$, the $4/5$ ($\mbox{mod } 1$)
resonance considered above is located at $\varepsilon_{res}
\approx 2.004 \cdot 10^{-216}$ ($m=0$), $\varepsilon_{res} \approx
2.670 \cdot 10^{-489}$ ($m=1$), $\varepsilon_{res} \approx 3.559
\cdot 10^{-762}$ ($m=2$), \dots. This is why it is impossible to
observe the resonant structure presented in Fig.~6a by means of
constructing sections of phase space of original system~(\ref{h}):
the relative magnitude of perturbation is too small. Such
situation is typical; this tentatively explains why the chaotic
layers in phase space sections of slowly perturbed Hamiltonian
systems usually do not show any sign of large inner regular
component (for examples see Ref.~\cite{NST97}): very fine tuning
of the values of the parameters is necessary to make the resonant
structure visible, and the resonant values of the relative
magnitude of perturbation for the most prominent resonances are
usually microscopic.

Finally, let us consider the influence of the magnitude of
perturbation on precision of our theoretical estimates as compared
to the numerical data. These data have been obtained by computing
the half-width $w_b$ of the chaotic layer of original
system~(\ref{h}) with $k=1$, $a=b$,  in dependence on the
magnitude of perturbation $\varepsilon$. Three values of
$\lambda$, namely, $\lambda = 0.01$, $0.1$, and $0.3$, have been
chosen, while the range of variation in $\varepsilon$ is six
orders of magnitude. The method utilizing Eq.~(\ref{wb1}) (the
``first method'', see above) has been employed for calculating
$w_b$. The computation results are shown in Fig.~7 as dots. Small
peaks ``perturbing'' the smooth curves are notable features of the
constructed dependences, especially in the case of the relatively
large value $\lambda = 0.3$. They are conditioned by the process
of encountering fractional resonances (such as the resonance the
phase portrait of which is presented in Fig.~6a), as $\varepsilon$
is varied.

The theoretical dependences given by Eq.~(\protect\ref{wb}) (for
$\lambda = 0.3$) and Eq.~(\protect\ref{wmr1}) are shown in Fig.~7
as solid curves. Eq.~(\protect\ref{wb}) is valid for description
of the case of $\lambda = 0.3$ because $0.79 < c < 4.92$ in the
given range of $\varepsilon$ (i.e., the $c$ values are far from
main resonance); and Eq.~(\protect\ref{wmr1}) is suitable for
$\lambda = 0.01$ because $0.06 < c < 0.20$, i.e., $c$ is close to
zero (main resonance). The theoretical curve for $\lambda = 0.1$
is not shown, because the numerical data for this case are very
close to that for the case of $\lambda = 0.01$. Theoretical
dependence~(\protect\ref{wmr1}) visually coincides with the
numerical data for the case of $\lambda = 0.01$ at $\log
\varepsilon < -2$.

The resonant peaks are not present in the theoretical curves, due
to the nature of approximation; so, the deviation from the
observed data increases locally at the resonances. Fig.~8
demonstrates the $\varepsilon$ dependence of the deviation $\delta
= (w_b^{num} - w_b^{theor})/w_b^{theor}$ of the observed data at
$\lambda = 0.01$ from theoretical dependence (\protect\ref{wmr1}).
At $\log \varepsilon < -3$ there is a plateau at the level $\log
\delta \approx -1.5$ with some resonant drops. Most probably its
emergence is due to the limit in precision of the numerical
determination of $w_b$ by the method used. The level $\log \delta
\approx -1.5$ corresponds to the accuracy of $\approx 3$\%.
Beginning at $\log \varepsilon \approx -3$, there is a linear rise
of $\log \delta$ with $\log \varepsilon$: the variation of $\log
\varepsilon$ by three orders of magnitude results in the variation
of $\log \delta$ by approximately one and a half orders of
magnitude, i.e., $\delta$ behaves like $\sim \varepsilon^{1/2}$.
This is in contrast with the expected precision of the separatrix
map~(\ref{sm}), which is $\sim \varepsilon$ (see
Section~\ref{mnr}). The different power law index is no wonder,
since the numerical data are not exact. However, detailed
understanding of the $\varepsilon$ dependence of the precision in
determining the chaotic layer width apparently needs further
numerical experiments and theoretical work.

\section{Conclusions}
\label{concl}

We have considered the problem of estimation of the width of a
chaotic layer near the separatrices of nonlinear resonance in a
Hamiltonian system. The model of a perturbed nonlinear pendulum,
representing, according to~\cite{C79}, a universal model of
nonlinear resonance, has been adopted for the analysis.

In a numerical experiment on determination of the width of the
chaotic layer, it has been shown that a sharp physical borderline
exists between slow and fast chaos at $\lambda \approx 1/2$
(Fig.~1).

A new method of description of the chaotic motion in the
neighbourhood of the separatrices of nonlinear resonance with
periodic perturbations of low frequency (the case of slow chaos)
has been presented. The method is based on the theory of
separatrix maps. Within the framework of the new method,
formulas~(\ref{yb}), (\ref{wb}) and Eqs.~(\ref{wmr}), (\ref{wmr1})
for the width of the chaotic layer have been obtained.

The developed theory has been compared to the results of numerical
experiments with original system~(\ref{h}); close agreement has
been observed at low and intermediate frequencies of perturbation
(Figs.~3, 5) and at small relative magnitudes of perturbation
(Figs.~7, 8). The phenomenon of bending of the chaotic
layer~\cite{S00} at low values of $\lambda$ has been directly
observed (Figs.~4, 5).

An important inference can be brought from the graphs in Figs.~3,
4, and 5: very sharp changes of the width of the chaotic layer are
possible upon only a small change in a parameter of the system.
However, fine tuning is necessary: large width of the layer is
reached inside narrow intervals of a parameter variation. Such
fine tuning may provide a potentially effective tool for
controlling the properties of the chaotic motion in Hamiltonian
systems.

The author is thankful to anonymous referees for useful remarks.
This work was supported by the Russian Foundation for Basic
Research (project \# 05-02-17555) and by the Programme of
Fundamental Research of the Russian Academy of Sciences
``Fundamental Problems in Nonlinear Dynamics''. The computations
were partially carried out at the St.\,Petersburg Branch of the
Joint Supercomputer Centre of the Russian Academy of Sciences.

\newpage

\begin{figure}
\centering
\includegraphics[width=0.75\textwidth]{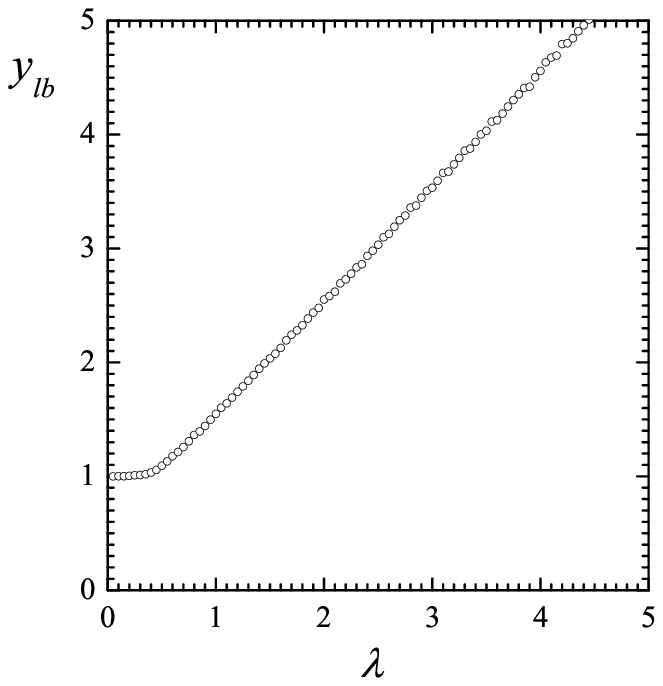}
\caption{The $\lambda$ dependence of the chaotic layer half-width
$y_{lb}$ in the case of the least perturbed border of the layer.}
\label{fig1}
\end{figure}

\begin{figure}
\centering
\includegraphics[width=0.75\textwidth]{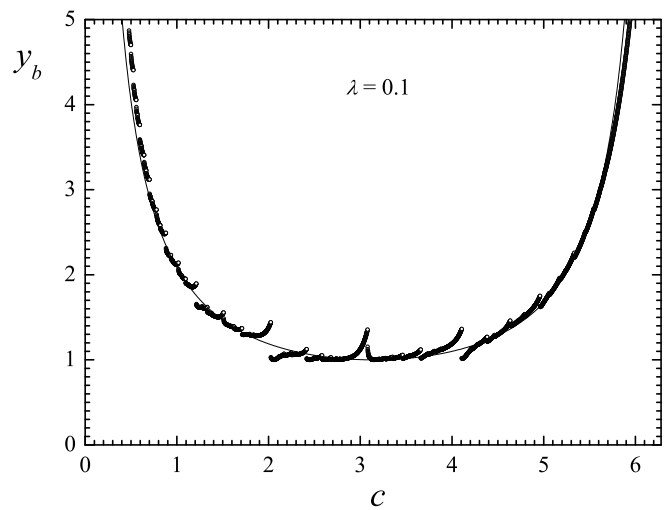}
\caption{The dependence of $y_b$ on $c$, computed (circles) and
theoretical (solid curve), $\lambda = 0.1$.}
\label{fig2}
\end{figure}

\begin{figure}
\centering
\includegraphics[width=0.75\textwidth]{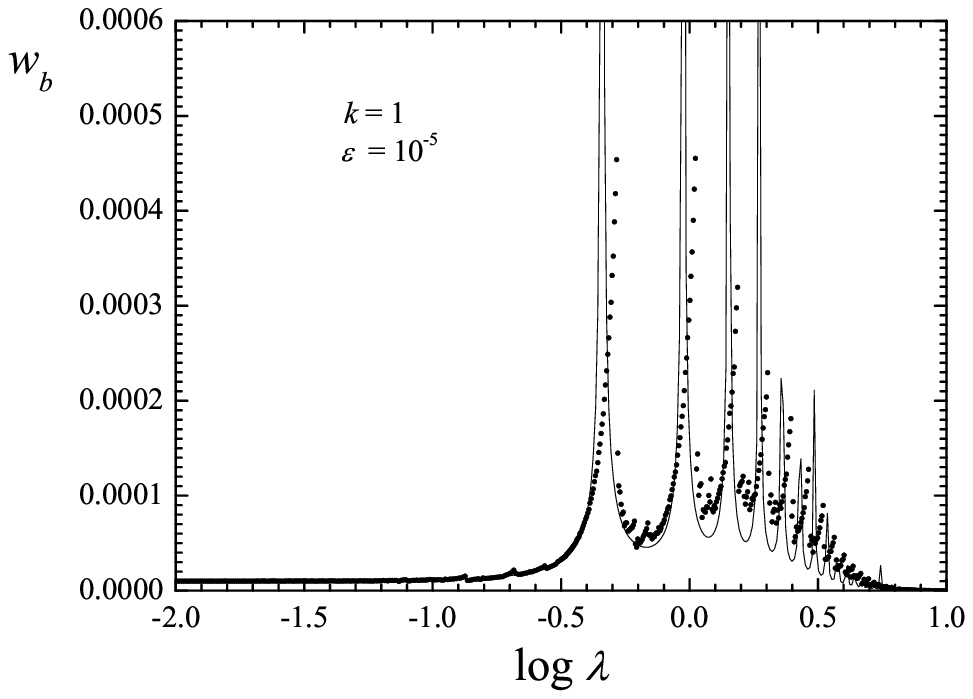}
\caption{The half-width $w_b$ of the chaotic layer of
system~(\protect\ref{h}) with $k=1$, $a=b$, in dependence on
$\lambda$: the results of direct computation by the first method
(dots) and the theoretical curve given by Eq.~(\protect\ref{wb}).}
\label{fig3}
\end{figure}

\begin{figure}
\centering
\includegraphics[width=0.75\textwidth]{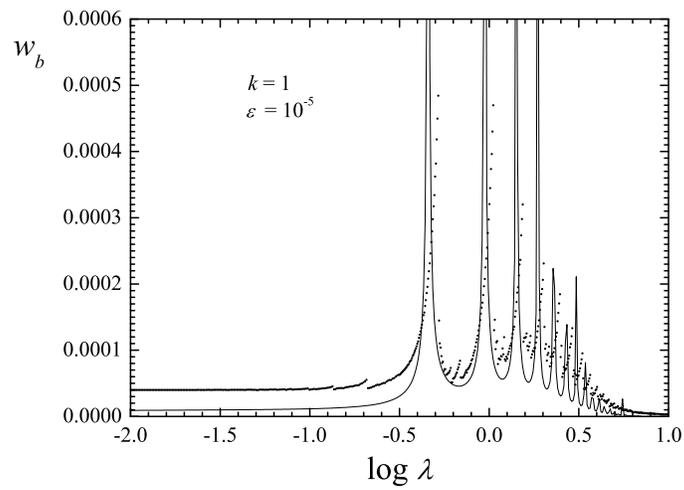}
\caption{The same as in Fig.~3, but the half-width is computed by
the second method (i.e., as the maximum energy deviation).}
\label{fig4}
\end{figure}

\begin{figure}
\centering
\includegraphics[width=0.75\textwidth]{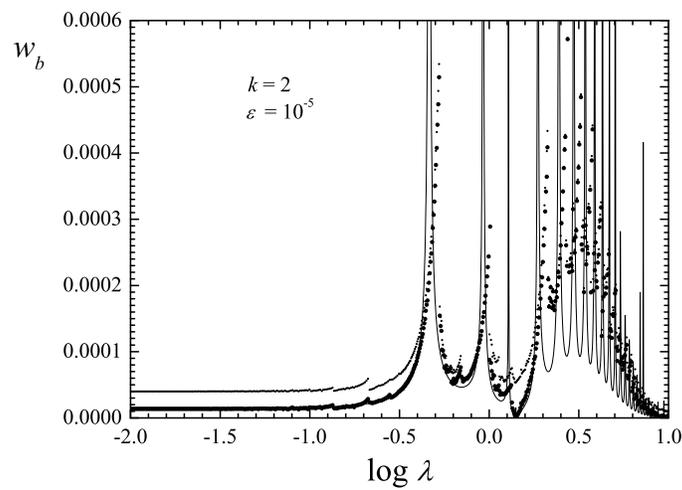}
\caption{The half-width $w_b$ of the chaotic layer of
system~(\protect\ref{h}) with $k=2$, $a=b$, in dependence on
$\lambda$: the results of direct computation by the first method
(big dots) and by the second method (small dots), and the
theoretical curve given by Eq.~(\protect\ref{wb}).}
\label{fig5}
\end{figure}

\begin{figure}
\centering
\includegraphics[width=0.75\textwidth]{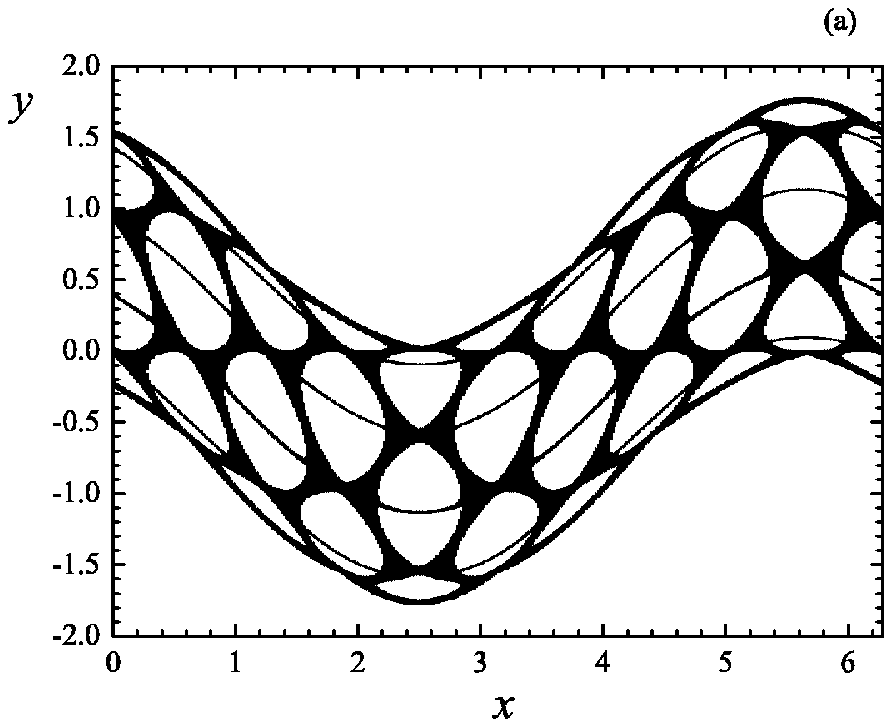}
\includegraphics[width=0.75\textwidth]{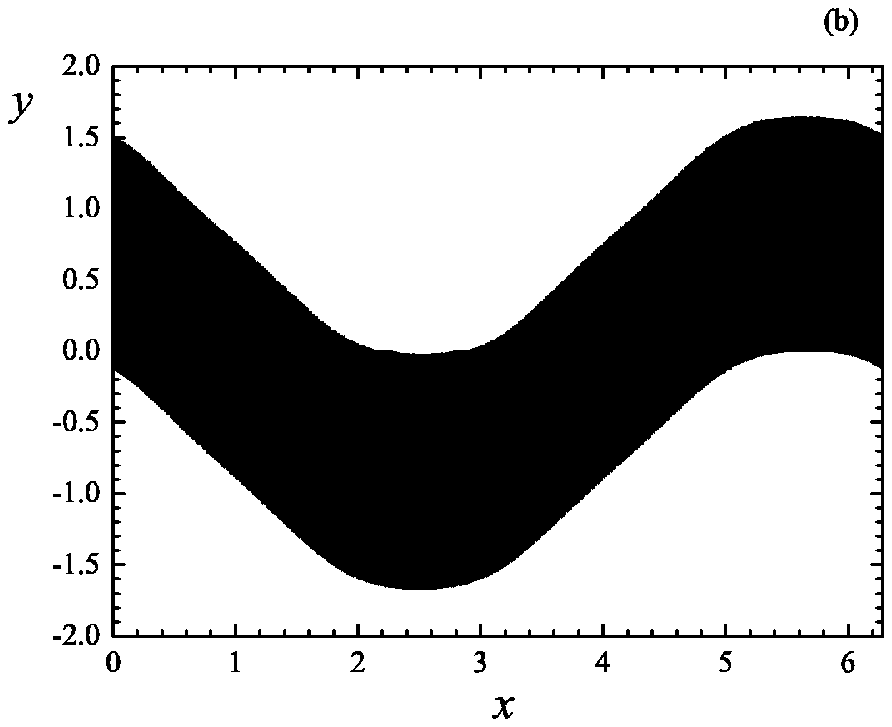}
\caption{The phase portrait of separatrix map~(\ref{sm1}), an
example. Only the chaotic component is shown. (a)~The case of the
$4/5$ resonance ($\lambda = 0.01$, $c = 5.0189$). (b)~The value of
the $c$ parameter is shifted away from the resonant case by $0.01$
(i.e., $c = 5.0289$).}
\label{fig6}
\end{figure}

\begin{figure}
\centering
\includegraphics[width=0.75\textwidth]{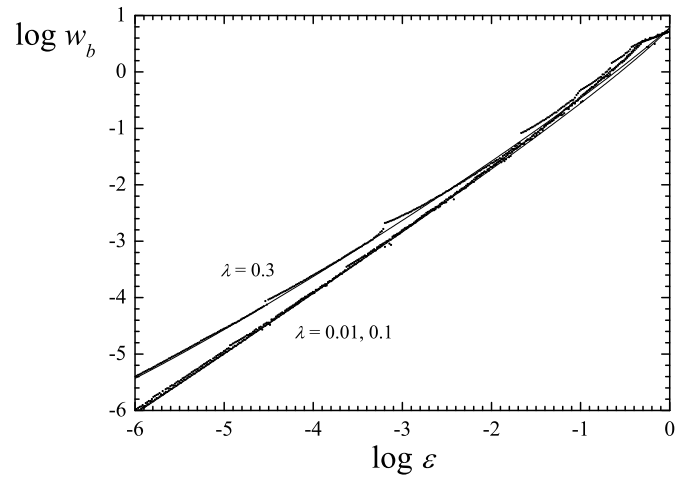}
\caption{The half-width $w_b$ of the chaotic layer of
system~(\protect\ref{h}) with $k=1$, $a=b$, in dependence on the
magnitude of perturbation $\varepsilon$: the results of direct
computation (dots) and the theoretical curves given by
Eq.~(\protect\ref{wb}) (for $\lambda = 0.3$) and
Eq.~(\protect\ref{wmr1}).}
\label{fig7}
\end{figure}

\begin{figure}
\centering
\includegraphics[width=0.75\textwidth]{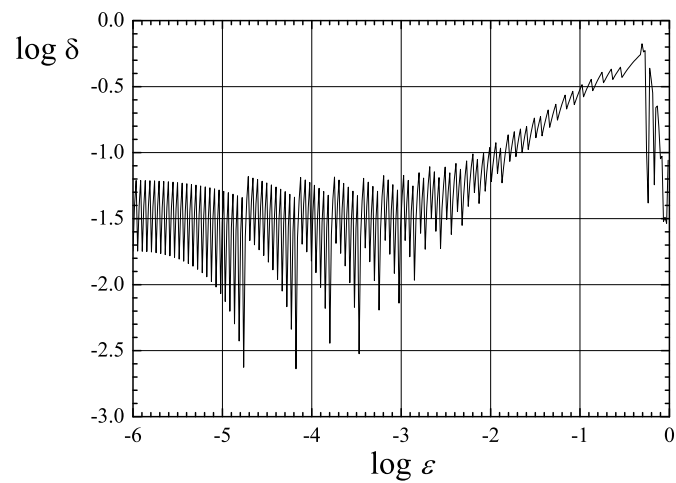}
\caption{The $\varepsilon$ dependence of the deviation $\delta$ of
the theoretical curve given by Eq.~(\protect\ref{wmr1}) from the
observed data, $\lambda = 0.01$.}
\label{fig8}
\end{figure}

\end{document}